\documentclass{iopart}

\usepackage{graphicx}
\usepackage{dcolumn}
\usepackage{bm}
\usepackage{nicefrac}
\usepackage{relsize}

\def\jpb{J. Phys. B }

\newcommand{\rtw}{\rightarrow}
\newcommand{\vect}[1]{\mathbf{#1}}
\newcommand{\cm}{cm$^{-1}$}
\newcommand{\GHzamu}{GHz$\cdot$amu}
\newcommand{\MHzfm}{MHz/fm$^2$}
\newcommand{\kmsec}{km\,s$^{-1}$}
\newcolumntype{b}{D{(}{\ (}{-1}}

\newcommand{\TiII}{Ti~II}
\newcommand{\knms}{\ensuremath{k_\textrm{\relsize{-1}{NMS}}}}
\newcommand{\ksms}{\ensuremath{k_\textrm{\relsize{-1}{SMS}}}}

\begin{document}

\title{Isotope shift calculations in \TiII}

\author{J. C. Berengut$^1$, V. V. Flambaum$^1$ and M. G. Kozlov$^2$}
\address{$^1$School of Physics, University of New South Wales, Sydney 2052, Australia}
\address{$^2$Petersburg Nuclear Physics Institute, Gatchina, 188300, Russia}

\date{20 June 2008}

\begin{abstract}

We present an accurate \emph{ab initio} method of calculating transition energies and isotope shifts in the $3d$-transition metals. It extends previous work that combines the configuration-interaction calculation with many-body perturbation theory by including the effective three-body interaction and modification of the energy denominator. We show that these effects are of importance in \TiII. The need to develop methods that can accurately calculate isotope shifts in $3d$-transition metals comes from studies of quasar absorption spectra that seek to measure possible variation of the fine-structure constant $\alpha$ over the lifetime of the Universe. Isotope shift can also be used to measure isotope abundances in gas clouds in the early Universe, which are needed in order to test models of chemical evolution.

\end{abstract}

\submitto{\jpb}

\maketitle

\section{Introduction}

This work is motivated by studies of quasar absorption systems that are designed to probe the fine-structure constant, $\alpha = e^2/\hbar c$, in the early Universe. By comparing the absorbed frequencies with terrestrial measurements one can deduce whether $\alpha$ has changed. Some studies report significant deviation from zero change (e.g.~\cite{webb99prl,webb01prl,murphy03mnras,murphy07prl,murphy08mnras}), while others do not (e.g.~\cite{quast04aap,srianand04prl,chand06aap,levshakov07aap}).

These studies use the ``many multiplet'' method \cite{dzuba99prl} where many transitions in many ions are used. This method offers an order-of-magnitude improvement in sensitivity over the previous ``alkali-doublet'' method, but introduces a potential systematic effect related to the isotope abundances of the absorbers: the isotope abundance ratios in the absorbing systems could be very different from those on earth. A ``conspiracy'' of several differences may provide an alternative explanation for the observed differences in spectra \cite{murphy01mnrasB,murphy03ass}.

We can resolve this problem by taking combinations of the transition frequencies that are insensitive to either $\alpha$-variation or isotopic abundances \cite{kozlov04pra}. We can then remove the systematic effects from the $\alpha$-variation studies and simultaneously measure isotope abundances in the gas clouds. The measured isotope abundances can then be used to test models of chemical evolution in the Universe. However, to do this type of analysis we must know both the relativistic shift and the isotope shift of each transition used in the analysis. While the relativistic shifts have been calculated for all important ions used in quasar absorption studies, there are still some gaps in the isotope shift calculations because these are generally more difficult. In particular, the isotope shifts of the important Fe~II transitions are not known.

Previously we have calculated the isotope shift in atoms and ions with one valence electron using many-body perturbation theory \cite{berengut03pra}, and for many-valence-electron atoms and ions using a combination of configuration interaction (CI) and many-body perturbation theory (MBPT) \cite{berengut05pra,berengut06pra}. The CI+MBPT method compares well with accurate multiconfiguration Hartree-Fock approaches (see, e.g. carbon calculations \cite{carlsson95jpb,jonsson96jpb} compared in \cite{berengut06pra}; Mg~I calculations \cite{jonsson99jpb} compared in \cite{berengut05pra}), however we believe that our method is more readily applicable to heavier ions, hence this study of \TiII. This ion has additional importance because it has been observed in quasar absorption spectra. Combined with accurate laboratory wavelengths \cite{aldenius06mnras} and our previous calculations of the relativistic shift \cite{berengut04praB}, \TiII\ may be a useful probe of $\alpha$-variation.

In this study we extend our previous work on calculation of energy and isotope shift using the CI+MBPT method by including the effective three-body second-order MBPT operator (\sref{sec:sigma3}) and varying the MBPT energy denominator (\sref{sec:energydenominator}). We show that these effects are of some importance in \TiII, and will probably prove to be of importance for all $3d$-transition metals. In \sref{sec:experiment} we extract isotope-shift constants from experiment in order to benchmark our theory. This experiment is of particular interest due to the lack of isotope-shift data for ionised $3d$ elements. Our final results are presented in \sref{sec:results}, along with predictions for astronomically relevant transitions. Atomic units ($\hbar = e = m_e = 1$) are used throughout this paper except where otherwise stated.

\section{Method}

The isotope shift of an atomic transition comes from two sources: the nuclear recoil (``mass shift''), and the finite size of the nuclear charge distribution (``field shift''). The mass shift is more important for light elements, while for heavy elements the field shift dominates. In the case of \TiII, the field shift is small (we estimate it in \sref{sec:field_shift}); this paper is concerned with the mass-shift contribution, which is more difficult to calculate.

Recoil of a nucleus of mass $M$ causes a level energy shift
\begin{equation}
\frac{\vect{p}_N^2}{2M}
     = \frac{1}{2M} \left( \sum_i \vect{p}_i \right)^2
     = \frac{1}{2M} \sum_i p_i^2 + \frac{1}{M} \sum_{i<j} \vect{p}_i \cdot \vect{p}_j\ . \label{eq:smsdef}
\end{equation}
The first term on the right hand side is known as the normal mass shift (NMS), while the second is the specific mass shift (SMS). We use the non-relativistic form of the mass-shift operator; relativistic corrections for optical transitions in light atoms are on the order of few percent and can be neglected \cite{korol07pra}. We calculate the frequency shift of a transition between two isotopes with mass number $A$ and $A'$ as
\begin{eqnarray}
\label{eq:is}
\delta \nu^{A', A} &=& \nu^{A'} - \nu^{A} \nonumber \\
    &=& \left( \knms + \ksms \right) \left( \frac{1}{A'} - \frac{1}{A} \right)
      + F \delta \langle r^2 \rangle ^{A', A} \ .
\end{eqnarray}
The normal-mass-shift constant is easily calculated from the transition frequency:
$\knms = -\nu/1823$,
where the value 1823 refers to the ratio of the atomic mass unit to the electron mass. The last term in this equation is the field-shift component which depends on the change in mean-square nuclear radius $\langle r^2 \rangle$ and the field-shift constant $F$, calculated in \sref{sec:field_shift}.

To calculate the specific-mass-shift constant, \ksms, we use the all-order finite-field scaling method. Here a rescaled two-body SMS operator is added to the Coulomb potential everywhere that it appears in an energy calculation:
\begin{equation}
\label{eq:tilde_Q}
\tilde{Q} = \frac{1}{\left| \vect{r}_1 - \vect{r}_2\right|} + \lambda \vect{p}_1 \cdot \vect{p}_2 \ .
\end{equation}
We recover the specific-mass-shift constant as
\begin{equation}
\ksms = \frac{d\omega}{d\lambda}\bigg{|}_{\lambda = 0}\ .
\end{equation}
The operator $\tilde{Q}$ has the same symmetry and structure as the Coulomb operator (see Appendix~A in Ref.~\cite{berengut06pra}).

\section{Energy Calculation}
\label{sec:energy}

To calculate energies, and hence transition frequencies, we use the CI+MBPT method \cite{dzuba96pra} implemented with the atomic structure package AMBiT. This package was previously used to calculate isotope shifts in Mg~I \cite{berengut05pra} and carbon ions \cite{berengut06pra}. It is presented in detail in \cite{berengut06pra}; here we will present only the salient points of the calculation, as well as some extensions that go beyond what was done in \cite{berengut06pra} (it should be noted, however, that these extensions have previously been proposed in \cite{dzuba96pra}).

The first step is to solve the Dirac-Fock equations for the core and valence electrons. From this we generate a single-particle basis set that includes the core and valence states and a large number of virtual states. Then we do the full configuration-interaction (CI) calculation in the frozen-core approximation. Here the many-electron wavefunction is expressed as a linear combination of Slater determinants $\left| I \right>$
\[
\psi = \sum C_I \left| I \right> \ ,
\]
where the $C_I$ are obtained from the eigenvalue problem
\begin{equation}
\label{eq:CI_matrix}
\sum_J H_{IJ} C_J = E C_I
\end{equation}
and $H$ is the CI Hamiltonian.

Core-valence correlation effects (that necessarily go beyond the frozen-core approximation) are included using an MBPT operator $\Sigma$, which is added to the CI Hamiltonian (see Section~III of \cite{berengut06pra}). In this paper we calculate $\Sigma$ to second order in the operator $\tilde{Q}$ (\Eref{eq:tilde_Q}), leading to the modified eigenvalue problem
\begin{equation}
\label{eq:CI_MBPT_matrix}
\sum_J \left( H_{IJ} + 
     \sum_M \frac{\left< I \right| H \left| M \right> \left< M \right| H \left| J \right>}{E - E_{M}}
     \right) C_J = E C_I \ .
\end{equation}
The states $\left| M \right>$ include all Slater determinants of the single-particle basis that have core excitations. We further separate the MBPT operator into one-valence-electron and two-valence-electron parts: $\Sigma^{(1)}$ and $\Sigma^{(2)}$, respectively. Goldstone diagrams and analytical expressions for these are given in \cite{berengut06pra}. All diagrams are included in the current work, including the box diagrams of $\Sigma^{(2)}$ which have the ``wrong'' multipolarity of the Coulomb interaction and were deliberately excluded from \cite{berengut06pra}.

\subsection{Effective three-body interaction: $\Sigma^{(3)}$}
\label{sec:sigma3}

\begin{figure}[htb]
\caption{Effective three-valence-electron interaction of $\Sigma$.}
\label{feyn:sigma3}
\centering
\includegraphics{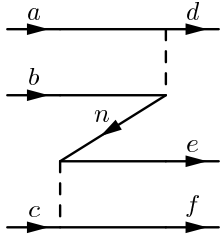}
\end{figure}

It is mentioned in \cite{berengut06pra} that there exists in the second order of perturbation theory an effective three-body interaction, where three valence electrons interact via the core, represented by a Goldstone diagram (\fref{feyn:sigma3}) with three external lines.

The diagrams of this type are quick to calculate: there is only one internal summation and no summation over virtual states. However the number of corresponding effective radial integrals is huge, and storage is not possible. Therefore we generate these diagrams as needed and seek a way of restricting them. In practice we only need to include orbitals from the leading configurations. For the \TiII\ transitions we're interested in this means that we only need to include $\Sigma^{(3)}$ in \Eref{eq:CI_MBPT_matrix} when either $\left| I \right>$ or $\left| J \right>$ represent $3d^2 4s$ or $3d^2 4p$ terms. In fact, the majority of the contribution comes from the case where both $\left| I \right>$ and $\left| J \right>$ represent leading configurations.

\subsection{Variation of energy denominator in perturbation theory}
\label{sec:energydenominator}

Another open question in the CI+MBPT method is how to deal with the energy denominator in \Eref{eq:CI_MBPT_matrix}. This question is discussed in detail in \cite{kozlov99os}; see also \cite{dzuba96pra,porsev01pra}. The basic problem is that we must make some approximation to the $E$ in the energy denominator in order to generate the modified matrix elements before solving the eigenvalue problem and obtaining the energy spectrum $E$. Two reasonable formalisms are the Rayleigh-Schr\"odinger (RS) and Brillouin-Wigner (BW) perturbation theories.

In \cite{berengut06pra} we used a BW method where all connected diagrams are evaluated at energies that correspond to the main configuration (i.e. the lowest valence energy). It is similar to taking an average (Dirac-Fock) energy of the states for $E$. In this paper we explore just one method of extending this, where we add a constant $\delta$ to every energy denominator:
\[
\frac{1}{E - E_M} \rtw \frac{1}{E - E_M + \delta}
\]
Our idea is to make the energy $E$ closer to the valence electron energy; to this effect we take $\delta = E^{CI} - E^{DF}$ for the ground-state energy. In effect this ``corrects'' $E$ for the ground state, replacing the Dirac-Fock value with the CI energy. In our case $\delta \approx -0.69$; we alter it depending on the values of $E^{CI}$ and $E^{DF}$ which in turn depend on the particulars of the calculation, and in particular on the SMS coefficient $\lambda$ (\Eref{eq:tilde_Q}). Note that this shift can be further justified on rather general grounds because it restores the correct asymptotic behaviour in the BW theory for a large number of particles \cite{kozlov99os}.

\section{Analysis of experiment}
\label{sec:experiment}

We compare our calculations to the Doppler-free spectroscopy measurements made by Gianfrani \etal\ \cite{gianfrani91oc}. They have made measurements of the $3d^24s\ ^4\!F_{9/2}$ -- $3d^24p\ ^4\!F_{9/2}^o$ and $3d^24s\ ^4\!F_{9/2}$ -- $3d^24p\ ^4\!G_{11/2}^o$ transitions for the isotope pairs 46-48 and 46-50. Their results are presented in table 1 of \cite{gianfrani91oc}; they record the following:

\begin{tabular}{lcb}
\multicolumn{1}{c}{Transition} & Isotope couple & \multicolumn{1}{c}{IS (MHz)}\\
\mr
$3d^24s\ ^4\!F_{9/2}$ -- $3d^24p\ ^4\!F_{9/2}^o$
    & 46-48 &  746(8)  \\
    & 46-50 & 1452(10) \\
$3d^24s\ ^4\!F_{9/2}$ -- $3d^24p\ ^4\!G_{11/2}^o$
    & 46-48 &  441(6) \\
    & 46-50 &  847(7) \\
\end{tabular}

However when one looks at the raw data for these measurements (Fig. 4 in \cite{gianfrani91oc}) it is clear that there is a misprint in the second transition: one may easily see that the $^4\!G_{11/2}^o$ isotope shift is only very slightly smaller than the $3d^24p\ ^4\!F_{9/2}^o$. In fact, one can take a rough reading from the fitted Lorentzian curves of Fig. 4a and obtain $\delta \nu^{50, 46} = 1425(150)$ MHz and $\delta \nu^{48, 46} = 725 (150)$ MHz.

Additionally, a mistake seems to have been made in the calculation of the normal mass shift. By taking the difference between the total isotope shift and residual shift in table 1 of \cite{gianfrani91oc}, one obtains 236.3 MHz and 453.7 MHz for the 46-48 and 46-50 isotope pairs, respectively, and independent of transition energy. Our calculated values of NMS are shown in \tref{tab:sms_experiment}.

Due to these considerations we have used the reported values of the measured isotope shift for the $3d^24s\ ^4\!F_{9/2}$ -- $3d^24p\ ^4\!F_{9/2}^o$ transition, and for the $3d^24s\ ^4\!F_{9/2}$ -- $3d^24p\ ^4\!G_{11/2}^o$ transition we use our own rough reading from the data with large errors.

\fulltable{\label{tab:sms_experiment}
Extraction of the specific mass shift constant, \ksms, from experiment. All isotope shift components are in MHz; \ksms\ is in \GHzamu. Our calculation of the field-shift (FS) contribution is presented in \sref{sec:field_shift}.}
\begin{tabular}{@{}lccbcbbb}
\br
\multicolumn{1}{c}{Transition} & $\lambda$ (\AA) & Isotopes & \multicolumn{1}{c}{IS (expt.)} & NMS & \multicolumn{1}{c}{FS} & \multicolumn{1}{c}{SMS} & \multicolumn{1}{c}{\ksms\ (\GHzamu)} \\
\mr
$3d^24s\ ^4\!F_{9/2}$ -- $3d^24p\ ^4\!F_{9/2}^o$ & 3235 &
    46 -- 48 &  746(8)  & 460 &  43(19) & 243(27) & -268(30) \\
 && 46 -- 50 & 1452(10) & 884 & 103(21) & 465(31) & -267(18) \\
$3d^24s\ ^4\!F_{9/2}$ -- $3d^24p\ ^4\!G_{11/2}^o$ & 3350 &
    46 -- 48 &  725(150) & 445 &  43(19) & 237(169) & -262(187) \\
 && 46 -- 50 & 1425(150) & 854 & 103(21) & 468(171) & -269(98) \\
\br
\end{tabular}
\endfulltable

In \tref{tab:sms_experiment} we extract \ksms\ from the experimental values of isotope shift. We remove the NMS and field shift (FS) components (the field shift is calculated in \sref{sec:field_shift}), and then remove the mass-dependence from the residual SMS. We conclude from \tref{tab:sms_experiment} that for consistency with experiment one requires values of the specific-mass-shift constant of $\ksms = -267\,(18)$ \GHzamu\ and $-269\,(98)$ \GHzamu\ for the $3d^24s\ ^4\!F_{9/2}$ -- $3d^24p\ ^4\!F_{9/2}^o$ and $3d^24s\ ^4\!F_{9/2}$ -- $3d^24p\ ^4\!G_{11/2}^o$ transitions, respectively.

\subsection{Field shift}
\label{sec:field_shift}

The field shift component of the isotope shift is given by
\[
\delta \nu^{A', A'} = F \delta \langle r^2 \rangle ^{A', A}
    = F \left( \langle r_{A'}^2 \rangle - \langle r_A^2 \rangle \right)
\]
where $\langle r^2 \rangle$ is the square-mean charge radius. We have used the \emph{rms} charge radii for \TiII\ isotopes tabulated in \cite{angeli98aph}, which result in $\delta \langle r^2 \rangle^{48, 46} = -0.104 (45)$ fm$^2$ and $\delta \langle r^2 \rangle^{50, 46} = -0.251 (45)$ fm$^2$.

We have developed two methods of calculating the field shift constant, $F$. The first is to simply vary the nuclear radius in the code (in AMBiT the nuclear charge has a fermi distribution) and calculate energy at each point. The field shift constant is then extracted as
\[
F = \frac{d\omega}{d\langle r^2 \rangle}\ .
\]

The second method is a scaling method where we take the difference in the potentials from two different nuclear charge radii and rescale it to increase the size of the effect:
$\delta U(r) = \lambda \left(\, U^{A'}(r) - U^{A}(r)\, \right)$,
where $U^{A}(r)$ is the nuclear potential of isotope $A$.
We add $\delta U(r)$ to the original nuclear potential, perform the energy calculation, and extract $F$ as
\[
F = \frac{1}{\delta \langle r^2 \rangle^{A', A}}\ \frac{d\omega}{d\lambda} \ .
\]
This method was used in \cite{berengut03pra} for single-valence-electron ions.

In practice, we have found that both methods give equivalent results within a few percent. Similarly, choice of basis sets makes little difference (our basis sets are discussed in the next section); any error in the FS constant is swamped by the experimental error in $\delta \langle r^2 \rangle$. We use $F \approx -410$ \MHzfm\ for both the $3d^24s\ ^4\!F_{9/2}$ -- $3d^24p\ ^4\!F_{9/2}^o$ and $3d^24s\ ^4\!F_{9/2}$ -- $3d^24p\ ^4\!G_{11/2}^o$ transitions. The spread of our results is less than 10 \MHzfm.

\section{Calculation and results}
\label{sec:results}

The ground state of \TiII\ has a $1s^2 2s^2 2p^6 3s^2 3p^6 3d^2 4s$ configuration, and we are interested in transitions to $3d^2 4p$ levels. A reasonable single-particle basis can be obtained by solving the self-consistent Dirac-Fock equations for the $1s^2 2s^2 2p^6 3s^2 3p^6 3d^2$ electrons (i.e. in the $V^{N-1}$ approximation) and generating other valence and virtual levels in the potential of these electrons.

Valence-valence correlations are included to all orders by the CI method. The $1s^2 2s^2 2p^6 3s^2 3p^6$ electrons are treated as a frozen core, and we include all single and double promotions from the leading configurations $3d^2 4s$ and $3d^2 4p$ in our calculation. Correlations with the frozen core (including the relatively important $3s$ and $3p$ orbitals) are treated using many-body perturbation theory, as explained in \sref{sec:energy}. Note that because the Dirac-Fock equations were not solved on the frozen core alone, the ``subtraction diagrams'' outlined in \cite{berengut06pra} must be included. There are two relevant basis sets: the MBPT basis that includes a very large number of virtual levels, and a more restricted CI basis that is a subset of the MBPT basis.

Our calculations are performed with a relativistic single-electron basis set made from B-splines, similar to those developed by Johnson \etal~\cite{johnson86prl}. This type of basis was shown to be effective in calculating transition frequencies and specific mass shifts in Mg~I and carbon ions \cite{berengut05pra,berengut06pra}.

We have included in the CI all single-particle levels up to $16spdf$ (that is, outside the frozen-core we have $3s$ -- $16s$, $3p$ -- $16p$, $3d$ -- $16d$, and $4f$ -- $16f$). This basis is large enough to effectively saturate the CI. For the sums over virtual levels in the MBPT diagrams we are able to include a much larger basis; we have used $33spdfg$. The results are presented in \tref{tab:bspline}, where we have separated the effects of the pure CI calculation from the effects of the MBPT operator. Note that the effective three-valence-electron part of the MBPT operator, $\Sigma^{(3)}$, has a non-negligible impact on the final results. This should be contrasted to other atoms (e.g. Tl \cite{dzuba96pra}, C \cite{berengut06pra}) where they were omitted. In our case this is not possible because of the large overlap between the $3s$ and $3p$ core orbitals and the $3d$ valence orbitals.

\Table{\label{tab:bspline} Frequencies and \ksms\ of \TiII\ levels relative to the ground state ($3d^24s\ ^4\!F_{3/2}$). Note that for this calculation we do not include modification of the energy denominator; i.e. $\delta = 0$. The last two lines show transitions where comparison with experiment is possible.}
\begin{tabular}{@{}lrrrrrr}
\br
\multicolumn{1}{c}{Level} & \multicolumn{1}{r}{CI} & \multicolumn{1}{r}{$\Sigma^{(1)}$} & \multicolumn{1}{r}{$\Sigma^{(2)}$} & \multicolumn{1}{r}{$\Sigma^{(3)}$} & \multicolumn{1}{r}{Total} & \multicolumn{1}{r}{Experiment} \\
\mr
\multicolumn{7}{c}{$\omega$ (\cm)} \\
$3d^24s\ ^4\!F_{9/2}$   &   331 &   -6 & 140 &  -4 &   461 &   393.44 \\
$3d^24p\ ^4\!G_{9/2}^o$ & 28418 & 1151 & 869 & 273 & 30711 & 29968.30 \\
$3d^24p\ ^4\!G_{11/2}^o$& 28668 & 1170 & 909 & 258 & 31005 & 30240.88 \\
$3d^24p\ ^4\!F_{9/2}^o$ & 29749 & 1226 & 902 & 188 & 32065 & 31301.01 \\
\\
\multicolumn{7}{c}{\ksms\ (\GHzamu)} \\
$3d^24s\ ^4\!F_{9/2}$   &   47.4 &  -12.2 &    1.0 &  -1.5 &   34.7 & \\
$3d^24p\ ^4\!G_{9/2}^o$ & -182.7 & -285.2 &  115.6 & -22.9 & -375.1 & \\
$3d^24p\ ^4\!G_{11/2}^o$& -161.4 & -290.2 &  118.4 & -26.9 & -360.2 & \\
$3d^24p\ ^4\!F_{9/2}^o$ &  -61.8 & -141.5 & -131.9 & -10.7 & -345.8 & \\
$^4\!F_{9/2}$ -- $^4\!G_{11/2}^o$ & & & & & -394.9 & \multicolumn{1}{b}{-269(98)} \\
$^4\!F_{9/2}$ -- $^4\!F_{9/2}^o$  & & & & & -380.5 & \multicolumn{1}{b}{-267(18)} \\
\br
\end{tabular}
\endTable

\Tref{tab:bspline_delta} presents a second set of results that include the modified energy denominator as discussed in \sref{sec:energydenominator}. As one might expect, the size of the MBPT contribution has been reduced because the energy denominators are now generally larger in magnitude. Nonetheless it is clear that the modification of energy denominators is a higher-order effect than the inclusion of $\Sigma$ itself.

\Table{\label{tab:bspline_delta} Frequencies and \ksms\ of \TiII\ levels relative to the ground state ($3d^24s\ ^4\!F_{3/2}$), including in $\Sigma$ our modification of the energy denominator of \Eref{eq:CI_MBPT_matrix}: $\delta = E^{CI}-E^{DF}$. The last two lines show transitions where comparison with experiment is possible.}
\begin{tabular}{@{}lrrrrrr}
\br
\multicolumn{1}{c}{Level} & \multicolumn{1}{r}{CI} & \multicolumn{1}{r}{$\Sigma^{(1)}$} & \multicolumn{1}{r}{$\Sigma^{(2)}$} & \multicolumn{1}{r}{$\Sigma^{(3)}$} & \multicolumn{1}{r}{Total} & \multicolumn{1}{r}{Experiment} \\
\mr
\multicolumn{7}{c}{$\omega$ (\cm)} \\
$3d^24s\ ^4\!F_{9/2}$   &   331 &  -7 & 109 &  -3 &   430 &   393.44 \\
$3d^24p\ ^4\!G_{9/2}^o$ & 28418 & 837 & 677 & 166 & 30099 & 29968.30 \\
$3d^24p\ ^4\!G_{11/2}^o$& 28668 & 850 & 709 & 159 & 30386 & 30240.88 \\
$3d^24p\ ^4\!F_{9/2}^o$ & 29749 & 969 & 625 & 109 & 31451 & 31301.01 \\
\\
\multicolumn{7}{c}{\ksms\ (\GHzamu)} \\
$3d^24s\ ^4\!F_{9/2}$   &   47.4 &   -9.8 &  1.3  &  -1.1 &   37.8 & \\
$3d^24p\ ^4\!G_{9/2}^o$ & -182.7 & -185.1 &  88.5 & -16.1 & -295.4 & \\
$3d^24p\ ^4\!G_{11/2}^o$& -161.4 & -189.3 &  90.5 & -18.4 & -278.6 & \\
$3d^24p\ ^4\!F_{9/2}^o$ &  -61.8 & -140.0 & -29.3 &  -8.4 & -239.5 & \\
$^4\!F_{9/2}$ -- $^4\!G_{11/2}^o$ & & & & & -316.4 & \multicolumn{1}{b}{-269(98)} \\
$^4\!F_{9/2}$ -- $^4\!F_{9/2}^o$  & & & & & -277.3 & \multicolumn{1}{b}{-267(18)} \\
\br
\end{tabular}
\endTable

In \tref{tab:astron} we present our calculations of isotope shift in the astronomically relevant transitions of \TiII, i.e. those seen in QSO spectra. The transition frequencies and SMS were calculated in the same manner as the transitions in \tref{tab:bspline_delta}, with $\delta = E^{CI}-E^{DF}$. The difference between these results and the $\delta = 0$ results was used to estimate accuracy. Field shift was calculated using the method of \sref{sec:field_shift}. The results are presented both in MHz and \kmsec: the latter is the preferred form for use in astronomy ($\delta \nu = \delta \lambda / \lambda \times c$ \kmsec).

\Table{\label{tab:astron} Frequencies and total isotope shifts of \TiII\ transitions to the ground state ($3d^24s\ ^4\!F_{3/2}$). Isotope shifts are relative to $^{48}$Ti.}
\begin{tabular}{@{}lrrrbbbb}
\br
\multicolumn{1}{c}{Transition} & $\lambda$ (\AA) & \multicolumn{2}{c}{$\omega$ (\cm)}
   & \multicolumn{2}{c}{$\delta\nu^{46, 48}$} & \multicolumn{2}{c}{$\delta\nu^{50, 48}$} \\
   & & \multicolumn{1}{r}{expt.} & \multicolumn{1}{r}{theory}
   & \multicolumn{1}{c}{(MHz)} & \multicolumn{1}{c}{(\kmsec)}
   & \multicolumn{1}{c}{(MHz)} & \multicolumn{1}{c}{(\kmsec)} \\
\mr
$3d^24p\ ^4\!G_{5/2}^o$ & 3385 & 29544 & 29636
    & -784(79)  & 0.265(27) & 742(73)  & -0.251(25) \\
$3d^24p\ ^4\!F_{3/2}^o$ & 3243 & 30837 & 30937
    & -756(98) & 0.245(32) & 717(90)  & -0.232(29) \\
$3d^24p\ ^4\!F_{5/2}^o$ & 3230 & 30959 & 31075
    & -742(95)  & 0.240(31) & 704(88)  & -0.228(28) \\
$3d^24p\ ^4\!D_{1/2}^o$ & 3074 & 32532 & 32732
    & -744(107) & 0.229(33) & 706(99)  & -0.217(30) \\
$3d^24p\ ^4\!D_{3/2}^o$ & 3067 & 32603 & 32821
    & -743(113) & 0.228(35) & 705(104) & -0.216(32) \\
\br
\end{tabular}
\endTable

\section{Conclusion}

In this paper we have calculated isotope shifts for \TiII\ as a test case for $3d$-transition metals because it has experimental data available for comparison. Our CI calculations for \TiII\ designate $3d$ as a valence orbital while keeping the $3s$ and $3p$ orbitals in the frozen core. This provides a great saving for the size of the CI calculation, but the $3s$ and $3p$ correlations are large, and must be included using MBPT in order to obtain good accuracy for the specific-mass shift. In this case we have shown that $\Sigma^{(3)}$ is important; we would expect it to be for all $3d$-transition metals.

We have also presented a modification of the MBPT energy denominator which consists of adding a constant, $\delta = E^{CI}-E^{DF}$, to the denominator. Here $E^{CI}$ and $E^{DF}$ are calculated for the ground state. This improves both the calculated frequencies and specific-mass-shift constants significantly. We have performed isotope-shift calculations for astronomically important transitions that are seen in quasar absorption spectra. Taken together with existing accurate laboratory wavelengths and relativistic-shift calculations, \TiII\ may be a useful probe of $\alpha$-variation.

Using the techniques presented in this paper we can perform accurate calculations of isotope shifts for other atoms of astronomical interest, including the $3d$-transition metals Fe~II, Cr~II, Ni~II, and Mn~II. These systems are very important in quasar absorption studies of $\alpha$-variation, yet their isotope shifts have not been measured. Furthermore, by comparing results of different methods, in particular the $\delta = 0$ and $\delta = E^{CI}-E^{DF}$ results, we can estimate the accuracy of our theoretical predictions.

\ack

This work was supported by the Australian Research Council. We are grateful to the APAC National Facility for providing computer time.

\section*{References}

\bibliographystyle{iopart-num}
\bibliography{references}

\end{document}